\documentclass[final,3p,11pt]{elsarticle}

\usepackage{cleveref}
\crefname{figure}{figure}{figures}
\usepackage{caption}
\usepackage{subcaption}
\usepackage[UKenglish]{babel}
\usepackage{alltt}
\usepackage{epstopdf}
\usepackage{setspace}

\usepackage{dcolumn}
\usepackage[perpage]{footmisc}
\usepackage{amsmath,amsfonts,amssymb,wasysym,ifsym}
\usepackage{color}
\usepackage{MnSymbol}
\usepackage{verbatim}
\usepackage{upgreek}
\usepackage{hyperref}
\usepackage{bm}
\usepackage{graphicx}

\newcommand \anu{${\bar \nu}_{\mathrm{e}}$}

\newcommand \gray{$\gamma$-ray}
\newcommand \lith{$^{6}$Li}

\newcommand \LiF{\lith F:ZnS(Ag)}

\journal{JINST}

\begin{document}

\begin{frontmatter}



\title{A novel segmented-scintillator antineutrino detector}


\makeatletter
\newcommand\footnoteref[1]{\protected@xdef\@thefnmark{\ref{#1}}\@footnotemark}
\makeatother

\address[a]{Universiteit Antwerpen, Antwerpen, Belgium}
\address[b]{University of Bristol, Bristol, UK}
\address[c]{Vrije Universiteit Brussel, Brussel, Belgium}
\address[d]{LPC Caen, ENSICAEN, Universit\'e de Caen, CNRS/IN2P3, F-14050 Caen, France}
\address[e]{CERN, 1211 Geneva 23, Switzerland}
\address[f]{Universiteit Gent, Gent, Belgium}
\address[g]{Imperial College London, Department of Physics, London, United Kingdom}
\address[h]{SUBATECH, CNRS/IN2P3, Universit\'e de Nantes, Ecole des Mines de Nantes, Nantes, France}
\address[i]{LAL, Univ Paris-Sud, CNRS/IN2P3, Universit\'e Paris-Saclay, Orsay, France}
\address[j]{University of Oxford, Oxford, UK}
\address[k]{SCK-CEN, Belgian Nuclear Research Centre, Mol, Belgium}
\address[l]{Center for Neutrino Physics, Virginia Tech, Blacksburg, Virginia, 24061, USA}
\address[m]{STFC, Rutherford Appleton Laboratory, Harwell Oxford, and Daresbury Laboratory, Warrington, United Kingdom}

\author[a]{Y.~Abreu \footnote{\label{ceaden} on leave from CEADEN, Havana, Cuba}}
\author[i]{Y.~Amhis}
\author[b]{L.~Arnold}
\author[d]{G.~Ban}
\author[a]{W.~Beaumont}
\author[i]{M.~Bongrand}
\author[i]{D.~Boursette}
\author[h]{J.~M.~Buhour}
\author[j]{B.~C.~Castle}
\author[b]{K.~Clark}
\author[k]{B.~Coup\'e}
\author[h]{A.~S.~Cucoanes \footnote{\label{roma} now at ELI-NP, Horia Hulubei National institute of Physics and Nuclear Engineering, Bucharest, Romania}}
\author[b]{D.~Cussans}
\author[a,e]{A.~De Roeck}
\author[c]{J.~D'Hondt}
\author[d]{D.~Durand}
\author[h]{M.~Fallot}
\author[h]{S.~Fresneau}
\author[k]{L.~Ghys}
\author[h]{L.~Giot}
\author[d]{B.~Guillon}
\author[h]{G.~Guilloux}
\author[g]{S.~Ihantola}
\author[a]{X.~Janssen}
\author[k]{S.~Kalcheva}
\author[c]{L.N.~Kalousis}
\author[k]{E.~Koonen}
\author[f]{M.~Labare}
\author[d]{G.~Lehaut}
\author[k]{J.~Mermans}
\author[f]{I.~Michiels}
\author[f,k]{C.~Moortgat}
\author[b,m]{D.~Newbold}
\author[l]{J.~Park}
\author[b]{K.~Petridis}
\author[a]{I.~Pi\~nera \footnoteref{ceaden}}
\author[b]{G.~Pommery}
\author[k]{L.~Popescu}
\author[h]{G.~Pronost}
\author[b]{J.~Rademacker}
\author[j]{A.~Reynolds}
\author[f]{D.~Ryckbosch}
\author[j]{N.~Ryder}
\author[b]{D.~Saunders}
\author[g]{Yu.~A.~Shitov}
\author[i]{M.-H.~Schune}
\author[j]{P.~R.~Scovell}
\author[i]{L.~Simard}
\author[g]{A.~Vacheret \footnote{\label{author} Corresponding author}} \ead{antonin.vacheret@imperial.ac.uk}
\author[k]{S.~Van Dyck}
\author[c]{P.~Van Mulders}
\author[a]{N.~van Remortel}
\author[a,c]{S.~Vercaemer} 
\author[j]{A.~Waldron}
\author[j,m]{A.~Weber}
\author[h]{F.~Yermia}

\begin{collab}
\centering (SoLid Collaboration)
\end{collab}

\date{\today}

\begin{abstract}
The next generation of very-short-baseline reactor experiments will require compact detectors operating at  surface level and close to a nuclear reactor. This paper presents a new detector concept based on a composite solid scintillator technology. The detector target uses cubes of polyvinyltoluene interleaved with $^6$LiF:ZnS(Ag) phosphor screens to detect the products of the inverse beta decay reaction. A multi-tonne detector system built from these individual cells can provide precise localisation of scintillation signals, making efficient use of the detector volume. Monte Carlo simulations indicate that a neutron capture efficiency of over 70~\% is achievable with a sufficient number of $^6$LiF:ZnS(Ag) screens per cube and that an appropriate segmentation enables a measurement of the positron energy which is not limited by \gray~leakage. First measurements of a single cell indicate that a very good neutron-gamma discrimination and high neutron detection efficiency can be obtained with adequate triggering techniques. The light yield from positron signals has been measured, showing that an energy resolution of 14\%/$\sqrt{E({\mathrm{MeV}})}$ is achievable with high uniformity. A preliminary neutrino signal analysis has been developed, using selection criteria for pulse shape, energy, time structure and energy spatial distribution and showing that an antineutrino efficiency of 40\% can be achieved. It also shows that the fine segmentation of the detector can be used to significantly decrease both correlated and accidental backgrounds.
\end{abstract}

\begin{keyword}
antineutrino \sep sterile neutrino \sep \lith F:ZnS
\end{keyword}

\end{frontmatter}


\section{Introduction}
\label{intro}

\subsection{Reactor \anu\ detection}
\label{reactor_exp}

The most widely used process to detect \anu originating from nuclear reactors is the inverse beta decay (IBD) reaction, which is the highest cross-section interaction process at typical reactor energies: 

\begin{equation}
	\label{eq:ibd} 
	\bar{\nu}_{e}+ p \rightarrow n + e^{+}  ~(E_{\bar{\nu}_{e}} > 1.805~\text{MeV}).
\end{equation}

Previous reactor neutrino experiments conducted at distances of less than 100\,m from the antineutrino source have used a variety of techniques to measure the rate and energy spectrum of \anu. Most approaches are sensitive to both the positron and the neutron in the final state, which are detected within a short time of each other. This coincidence technique, pioneered by Cowan and Reines~\cite{CowanReines}, provides a robust signature and reduces the chance of misidentifying background radiation as neutrino interactions.

The highly efficient $^{3}$He counter, widely available in the past, was an obvious choice for detecting neutrons~\cite{Knoll, Mills}. It was typically deployed in conjunction with a large active mass of liquid scintillator (LS), enclosed in large containers, that provided the \anu\ target. Detectors were composed of alternating layers of neutron counters and target sections read out by photomultiplier tubes (PMT). In most experiments, the rate of IBD coincidences was recorded but the Krasnoyarsk~\cite{krasnoyarskI, krasnoyarskII}, Rovno~\cite{Rovno88, Rovno91} and Bugey~\cite{Bugey1994, Bugey1995} experiments detected \anu\ using inert target volumes of water or polyethylene and counted interactions based on the excess rate of neutron captures in $^{3}$He. This simple technique made the measurement of the inverse beta decay rate possible to a precision of 2\%.

More recent experiments have moved from segmented to homogeneous volumes of liquid scintillators~\cite{Eguchi:2002dm}, doped LS+Li~\cite{Aleksan:1988qh} and LS+Gd~\cite{Apollonio:1997xe} to scale up the \anu\ target volume to several tonnes, allowing measurements at kilometre-scale distance. The large volume of LS enables a good containment of the visible energy, a necessity for measuring the antineutrino energy spectrum with good resolution. For these detectors, located underground, recent analyses have also shown that high precision oscillation measurements can be achieved using neutron capture on hydrogen~\cite{Abe:2013sxa, An:2016bvr}.

\subsection{Short-distance \anu\ detection}
\label{challenge_surface}

The re-evaluation of reactor \anu\ flux calculations that led to the so-called reactor anomaly~\cite{PhysRevC.83.054615,PhysRevD.83.073006,PhysRevC.84.024617} and the recent evidence for a deviation in the \anu\ energy spectrum shape, observed by three short baseline experiments ($\sim$ 1-2 km)~\cite{Abe:2014bwa,An:2015nua,RENO:2015ksa}, has put our understanding of the reactor flux model into question. New measurements at reactors, at closer distances than ever before, are needed to test possible explanations. The use of a highly segmented detector target to probe small length scale oscillations is mandatory, as is the ability to measure the \anu\ energy spectrum with high precision, above ground and at a distance from the reactor core of around 10\,m. 

The challenges posed by operating a detector under these conditions include:
  
\begin{enumerate}

	\item Cosmic-ray background radiation (neutrons, muons and hadrons) produced in the atmosphere interacts in the detector orders of magnitude more frequently than the expected \anu\ rate. The neutron and muon components are highly penetrating and can induce secondary reactions such as spallation and activation, which result in time-correlated backgrounds that are difficult to eliminate. Current experiments operate underground with an overburden of at least 50\,m water equivalent, which is not possible at most research reactor facilities. 

	\item High power research reactors are intense neutron and \gray~ sources, as this is usually their primary purpose. These emissions scale with the antineutrino signal and can thus be difficult to control, especially very close to the reactor core. The choice of experiment site, therefore requires careful assessment and control of reactor background conditions.  

	\item The high level of safety and security at a reactor impacts the choice of the detector components, the nature of the installation and the operational regime of the experiment.

\end{enumerate}

Both the anomalous rate deficit and the distortion of the energy spectrum are deviations of 5 to 10\% in amplitude and require therefore a detector capable of determining the antineutrino energy to a few percent with a reconstruction that has minimum dependence on the visible energy and the segmentation. 
Variations of detector response between cells after correction needs also to be at the percent level. 
Going further, an absolute measurement could greatly improve the constraints on both of these questions. 
At research reactor the current limit on absolute measurement is the reactor power measurement which is of the order of 3-5\%. 
An absolute neutron efficiency of a few percent or less would provide sufficient accuracy for an absolute measurement to be made. 
As large statistical samples are needed, an IBD efficiency above 30\% would ensure that enough IBD events are detected per reactor cycle and per year.

In the following sections we describe a new detector design that revisits the concept of a segmented detector and addresses the key challenges for a high precision \anu\ detection at short distance from a nuclear reactor. The detector technology is based on a combination of plastic and inorganic scintillators with a segmentation and geometry that provides a high neutron detection efficiency, an accurate energy measurement and maximises the signal-to-noise ratio for IBD events.

This detector technology will be used in a next generation very short baseline experiment called SoLid (Search for Oscillations with a \lith~detector). The detector will be used to probe possible new neutrino oscillation signals as a function of energy and distance from the source. The reduction of detector size and complexity compared to current \anu\ detectors also makes this technology an attractive option for future applications in reactor monitoring.

\section{Detector concept}
\label{solid_det_concept}

\begin{figure}[ht]
\begin{center}
\includegraphics[width=0.8\textwidth]{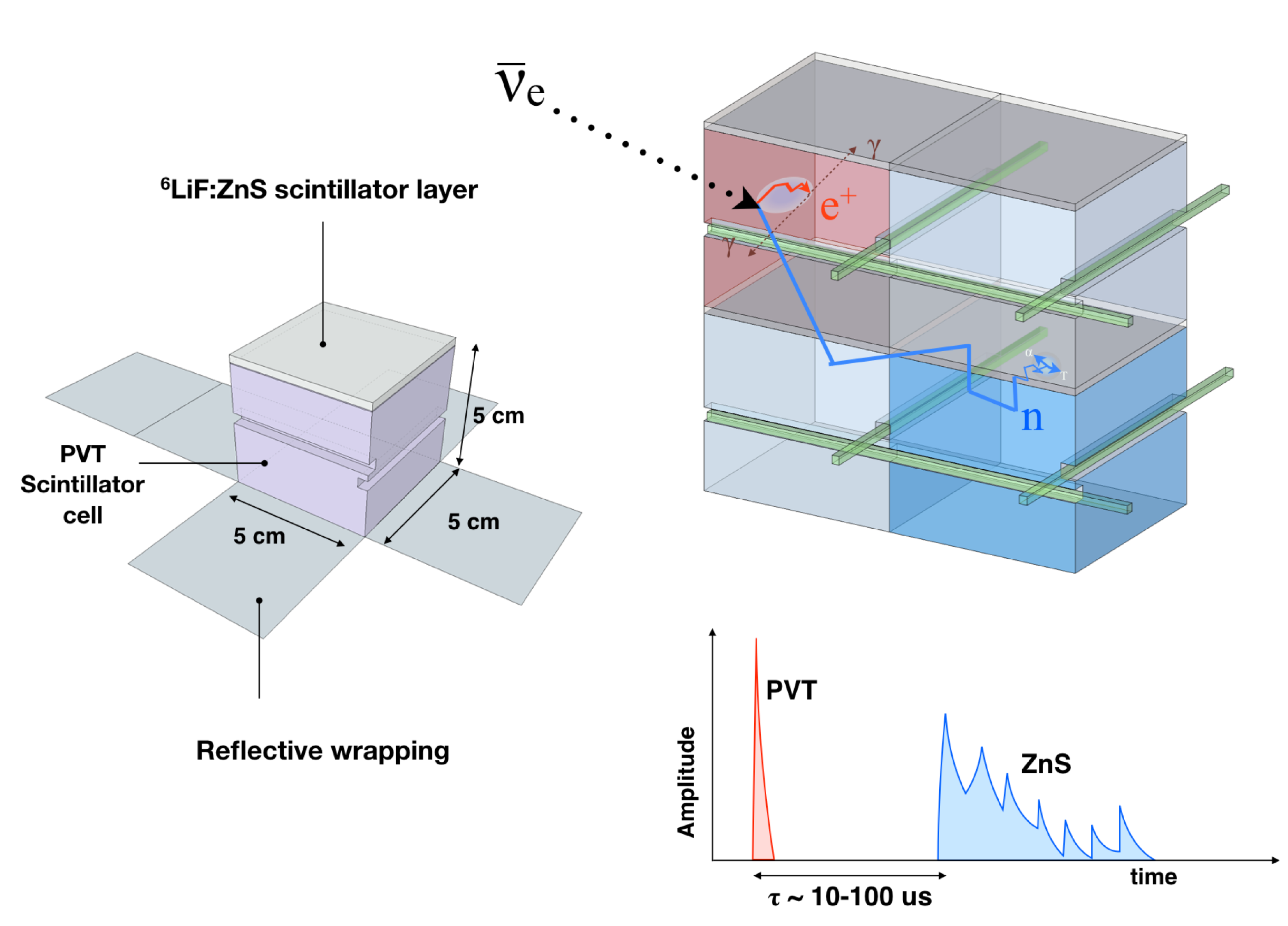}
\caption{\label{fig:detprincip} (Left) The PVT cube is covered with a layer of $^6$LiF:ZnS(Ag) and the assembly wrapped in a reflective material. (Top right) Principle of \anu\ detection in a volume made of separated voxels: wavelength shifting fibres placed in perpendicular orientations are used to collect the scintillation light from each cell of the array. (Bottom right) Time spectrum of PVT and ZnS(Ag) scintillation, used to identify the IBD reaction products.} 
\end{center}
\end{figure}

The basic detector cell consists of a cube of polyvinyltoluene (PVT) of ($5\times5\times5$)\,cm$^3$ in dimension, of which one or more faces are covered with thin sheets of $^6$LiF:ZnS(Ag) to capture and detect the neutron. A schematic drawing of a cell is shown in Fig.~\ref{fig:detprincip}. PVT is a mouldable plastic scintillator with high light yield and excellent optical transparency. Here it also acts as an effective moderator for neutrons. The cells are stacked in planar frames with a typical surface area of O(1m$^2$)and are arranged normal to detector-reactor axis. Several of these frames may be grouped to provide the required target volume and instrumented baseline.

The PVT provides a proton-rich \anu\ target and produces a prompt and rapidly decaying scintillation signal due to the positron from the IBD reaction. The neutron, produced at the same time, reaches thermal energies after scattering in the PVT over a period of up to hundreds of microseconds and is subsequently captured in the $^6$LiF:ZnS(Ag) layer. The high neutron capture cross-section of $^6$Li ensures that only a small fraction of neutrons are captured on hydrogen. The scintillation signal of ZnS(Ag) is significantly longer in duration than that of PVT and this difference is used to identify neutron capture, as shown in Fig.~\ref{fig:detprincip}. The IBD reaction products can thus be distinguished using pulse shape information and the time difference between the positron and neutron induced signals used for IBD selection against background. Due to the fine segmentation of the detector, the spatial distribution of scintillation signals can also be exploited for background reduction. This is discussed in later sections of this paper.

The scintillation light produced by both ZnS(Ag) and the PVT has wavelengths in the blue visible spectrum centred around 420 nm. The light is captured due to internal reflection at the PVT-air interface. The cube, along with its $^6$LiF:ZnS(Ag) layers, is wrapped in a reflective woven polymer fabric to further enhance light capture and to avoid leakage of light to neighbouring cells. The coupling between the neutron layer and the PVT cube, via a thin layer of air, also allows the photons produced in the ZnS(Ag) to be captured. Some of the scintillation light is subsequently trapped in wavelength shifting (WLS) fibres and re-emitted as green optical photons. The fibres are of the BC-91A type from St Gobain with a square cross section of $(3\times3)$\,mm$^2$ and lie within 5\,mm grooves in the faces of each cube, crossing each  detector plane in perpendicular directions. Placing the fibres onto two orthogonal faces of the cube enables the localisation of the scintillation signal via a hodoscope technique.

The wavelength-shifted light travels via the fibres to a silicon photomultiplier (SiPM) from Hamamatsu called MPPC model S12572-050P, with a surface area matching that of the fibre. Each fibre has a SiPM which is optically coupled to one of the end of each fibre, with a mirror attached to the other end. This technique has been successfully used in large detectors built for long baseline neutrino experiments~\cite{Abe2011106,Michael2008190} and has been proven to yield a uniform and robust light transmission. The SiPM pulse amplitude spectrum has a quantised form, corresponding to an integer number of single pixel avalanches (PA) triggered by the incoming photons. 

\begin{figure}[ht]
\begin{center}
\includegraphics[width=0.6\textwidth]{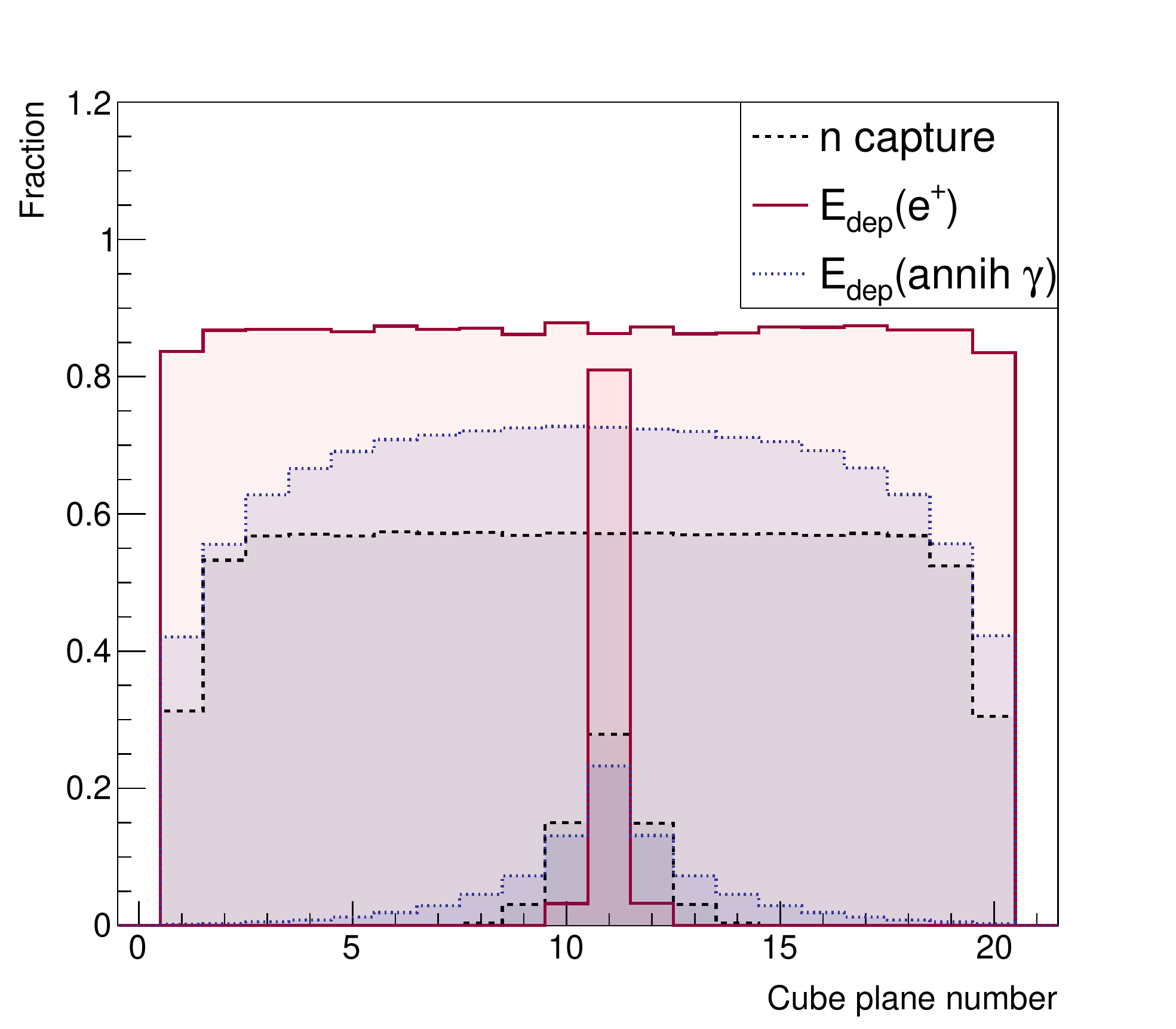}
\caption{\label{fig:detcontainment} The distribution of relative efficiency of neutron capture and energy deposition of the positron and the annihilation \gray shown across the detector volume for each detector plane and for one detector plane at position 11 close the middle of the volume.}
\end{center}
\end{figure}

The components described above allow for a relatively inexpensive and modular tonne-scale detector system that provides adequate containment of the neutron capture scintillation signal -- around a hundred times better than a LS + Gd detector, in terms of the volume needed to contain the scintillation signal -- and a robust and precise three-dimensional positioning of both positrons and neutrons. To illustrate these capabilities a large sample of IBD interactions was simulated in a 1\,m$^3$ SoLid detector, comprising 20 detection planes with a surface of 1~m$^2$ each. The G{\sc eant4}~\cite{Agostinelli:2002hh} package was used for simulation. Fig.~\ref{fig:detcontainment} shows three distributions: the fraction of the energy deposited by the positron, the fraction of energy deposited by the annihilation \gray~ and the neutron capture efficiency across the 20 detector planes. The distributions are shown both for IBD event vertices uniformly distributed over the detector volume and for IBD vertices confined to the centre plane of the detector (plane number 11). 
The simulation of neutron capture and transport was validated using the MCNP framework~\cite{mcnp5} and agreement with the G{\sc eant4} predictions was observed at a level of better than 5\%. 

The first set of distributions shows that the efficiency of capture of the neutron and detection of the positron is uniform throughout the detector, except for the planes located at the very edge. The 511 keV \gray~ produced by the annihilation of the positron are not well contained due to their large range (approximately 30 cm) in PVT. The second set of curves indicates that an IBD event is spatially very well contained within the detector: the positron will deposit most of its energy in the cell where the IBD reaction took place, whilst the neutron is very likely to be detected in a neighbouring cell. This spatial containment of the IBD event can be exploited to reduce time-correlated backgrounds and to obtain a statistical measure of the direction of the \anu. The containment of the products of the IBD reaction in a small number of cells maximises the usable volume provided by the detector. Because the annihilation \gray~ have a low energy and a weaker correlation with the IBD interaction point, they can not be so easily used in the reconstruction of the positron energy. However these characteristics may be used to help identify the positron signal.   

An additional layer of neutron moderator around the detector volume may be added to improve the homogeneity of the neutron capture efficiency across the detector, at the cost of a minimal increase in size. Simulation results show that the addition of a 5\,cm polyethylene neutron reflector, surrounding the whole detector volume, improves the overall neutron capture efficiency by 7\% and by 30\% in the layers closest to the edge of the detector.

\section{Neutron detection}
\label{n_det}

\subsection{Neutron identification}
\label{n_id}
 
Neutrons are detected using micro-composite \lith F:ZnS(Ag) `ND screen' scintillator screens from Scintacor~\cite{scintacor}. The neutron capture reaction on \lith{} is : 

\begin{equation}
	n + {}^{6}\mathrm{Li}  \rightarrow {}^{3}\mathrm{H}+ \alpha + 4.78~\mathrm{MeV}.
\end{equation}

\begin{figure}[ht]
\begin{center}
\begin{subfigure}{1.0\linewidth}
  \centering
  \includegraphics[width=.5\linewidth]{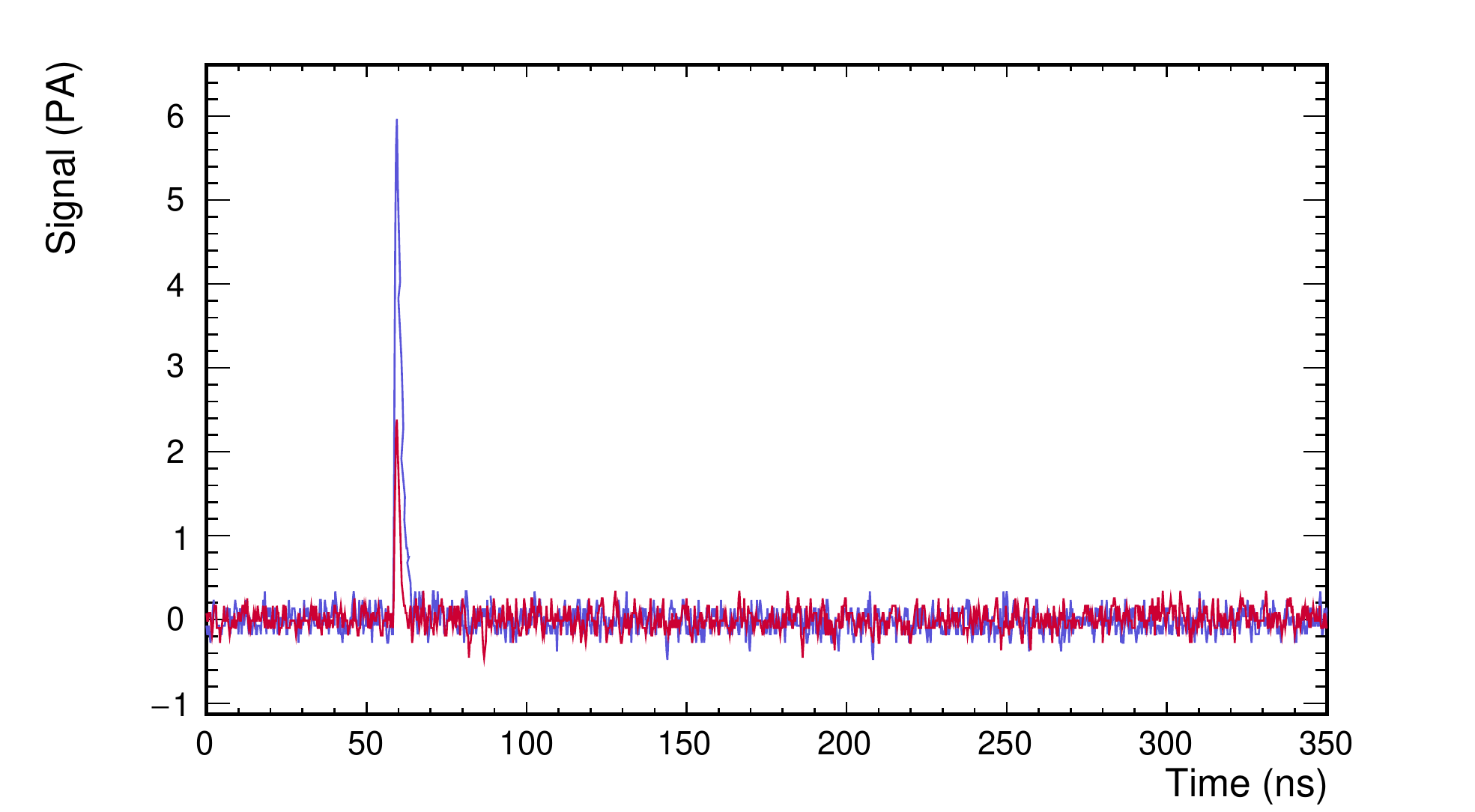}
  \label{fig:waveformEm}
\end{subfigure}
\begin{subfigure}{1.0\linewidth}
  \centering
  \includegraphics[width=.5\linewidth]{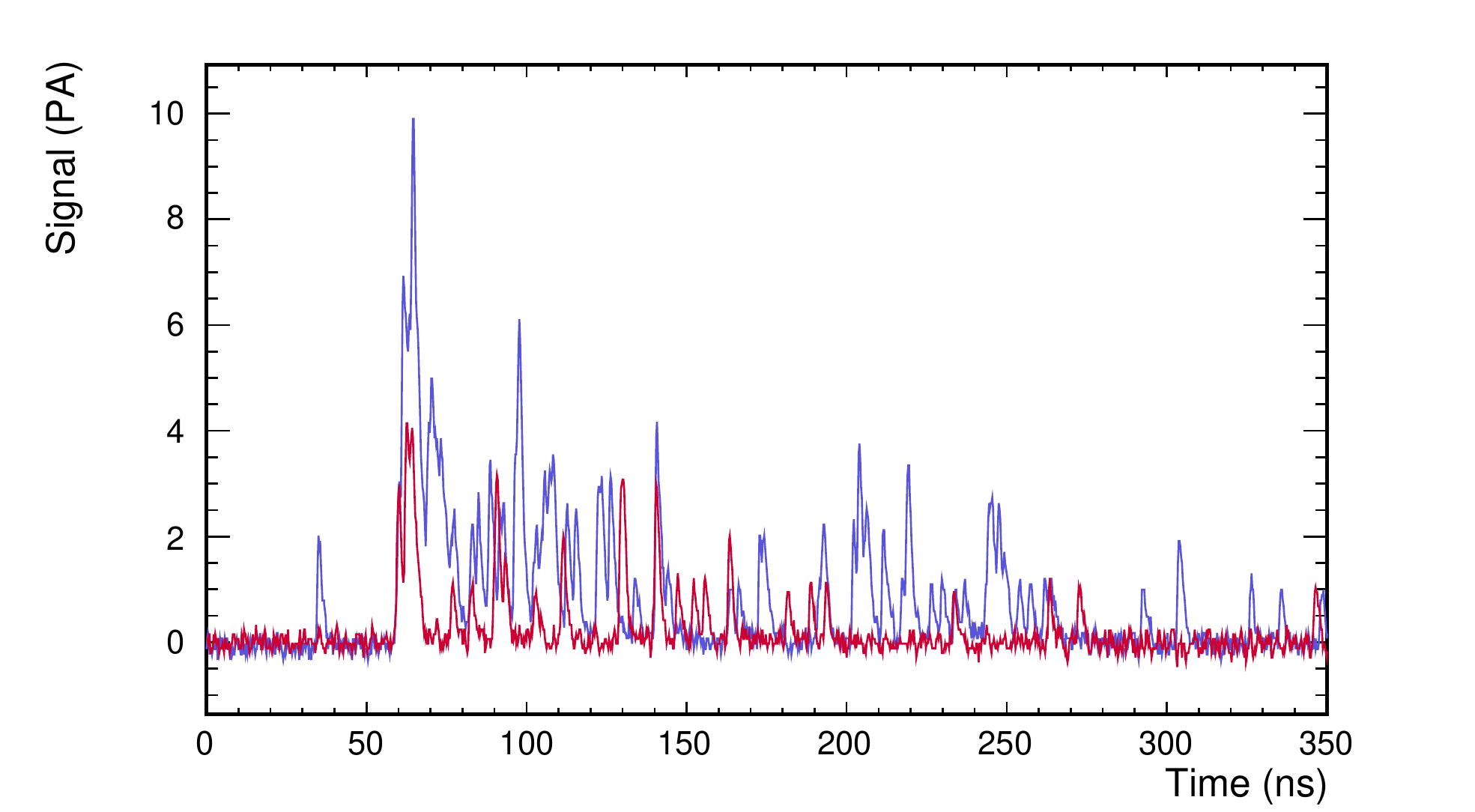}
\end{subfigure}
\caption{Typical charged particle signal in ZnS or PVT (top) and a neutron signal waveform occuring in the \lith F:ZnS(Ag) sheet (bottom). The blue and red curves represent the waveforms detected by SiPMs connected to two different optical fibres crossing a single cell in perpendicular directions. \label{fig:waveform} 
\label{fig:waveformNeutron}}
\end{center}
\end{figure}

This reaction has a high capture cross-section for thermal neutrons (936\,barns compared to 0.33\,barns for hydrogen). The large number of~\lith~atoms present in the screen results in a substantial probability of capture per neutron crossing. The \lith~ decay products have sufficient kinetic energy to travel a few tens of microns in the mixture and excite surrounding ZnS grains. As the quenching factor of the inorganic ZnS(Ag) scintillator is small, discrimination between neutron capture events and excitation by \gray~ and charged particles is possible. The high density of ionisation energy following neutron capture leads to a large population of excited states, with a wide range of recombination transitions to the ground state.  The resulting signal exhibits a sharp rising edge and a relatively long tail lasting several microseconds. Excitation by lighter and less intensely ionising particles results in a lower energy density, populating only states with lower excitation energy, which have a shorter decay time of less than 100\,ns, similar to the PVT signal. Typical examples of overall scintillation response to charged particles and neutrons are shown in Fig.~\ref{fig:waveform}, recorded in a single detector cell exposed to a moderated AmBe source and read out by two fibres. 

\begin{figure}[ht]
\begin{center}
\includegraphics[width=.7\textwidth]{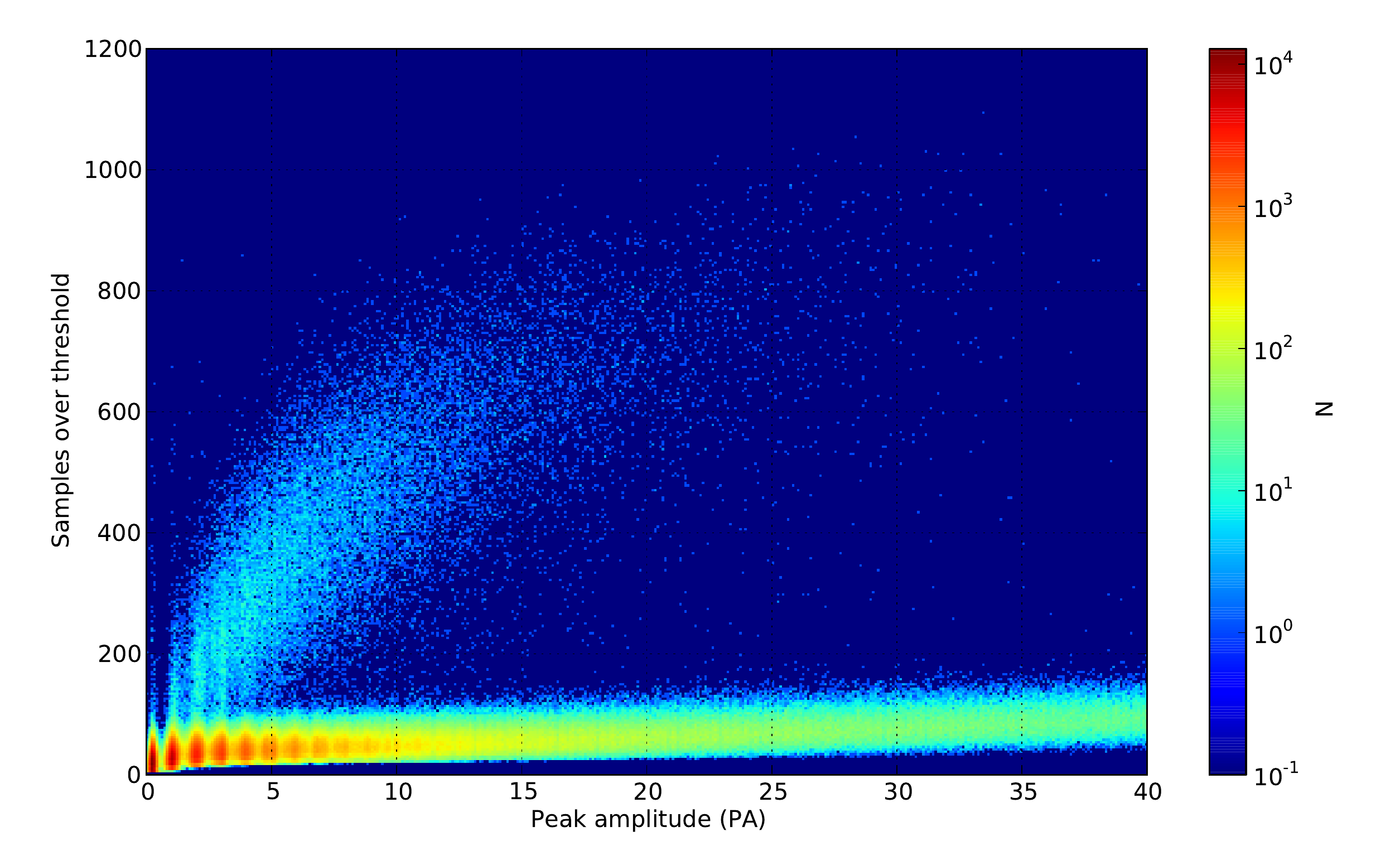}
\caption{Demonstration of a pulse discrimination method based on the number of samples above a chosen threshold and the pulse amplitude for a detector cell exposed to an AmBe source. The horizontal band corresponds to PVT and ZnS signals produced by recoil protons and \gray. Neutron events can be seen on the upper part of the figure. \label{fig:neutronPID}}
\end{center}
\end{figure}

An example of pulse shape discrimination based on time over threshold is illustrated in Fig.~\ref{fig:neutronPID}. The fast signals are located in the horizontal band at the bottom and the neutron capture population is in the band with a higher number of samples over threshold. The saturation seen at high value of time over threshold is due to the fixed sampling duration of the acquisition system, which was around 5 microseconds and smaller than the long decay time component of the ZnS.  

A consequence of the large number of continuously-sampled channels for a massive fine granularity detector is an extreme high recorded data rate; for the detector geometry described earlier, around 0.5\,Tbit/s (for a desired sampling rate of 40 MHz). Selective and robust neutron pulse recognition techniques (e.g.~\cite{adams1978, gamage2011}) must therefore be implemented in a real-time trigger system based on Field Programmable Gate Arrays.
 
\subsection{Neutron capture efficiency}
\label{n_cap}

The theoretical capture efficiency for \lith F:ZnS(Ag) screens is a function of their thickness and \lith F to ZnS ratio. A typical layer with a molecular ratio of 1:2 and a thickness of 250\,$\upmu$m has a probability of capture of around 26\% per neutron crossing according to manufacturer data. In the proposed detector design the IBD neutron may cross the same screen more than once through multiple scatters on protons, which raises the IBD neutron capture efficiency. The neutrons may also be captured by the hydrogen of the PVT or escape from the detector through the air gaps around the fibres. 
Neutron capture efficiency may be increased by the addition of multiple screens around the faces of the PVT cubes. The increase of \lith\ content per voxel also reduces the mean capture time. Table~\ref{table_nScreens} shows the simulated \lith\ capture efficiency (both on lithium-6 and hydrogen) and mean capture time on lithium-6 for IBD neutrons in configurations with a different number of neutron screens per cube.

\begin{table}[htp]
\caption{Impact of number of \lith F:ZnS(Ag) sheets on the neutron capture efficiency on lithium-6, mean capture time on lithium-6 and efficiency of capture on hydrogen. \label{table_nScreens}}
\begin{center}
\begin{tabular}{|c || c || c || c |}
\hline 
Number of  & \lith\ capture  &  Mean capture & n capture on H\\
\lith F:ZnS(Ag) sheets & eff. & time ($\mu$s) & eff. \\
\hline 
\hline 
1 	&    0.51 & 102 & 0.28  \\ 
2 	&    0.66 &  65  &  0.17  \\
3 	&    0.73 &  49  &  0.12  \\
\hline 

\end{tabular}
\end{center}
\label{tab:ncapture}
\end{table}%

\subsection{Neutron capture uniformity}
\label{n_uni}

The area of neutron absorber screen required to equip a tonne-scale detector is of the order of a few tens of square metres. In order to achieve the required few percent accuracy in neutron detection efficiency calibration, it is necessary to have a highly uniform capture probability throughout the detector. \lith\ content and screen thickness are important parameters affecting this uniformity and the effect of variation in these parameters was studied with a G{\sc eant4} detector simulation.

\begin{figure}[ht]
\begin{center}
\includegraphics[width=0.5\textwidth]{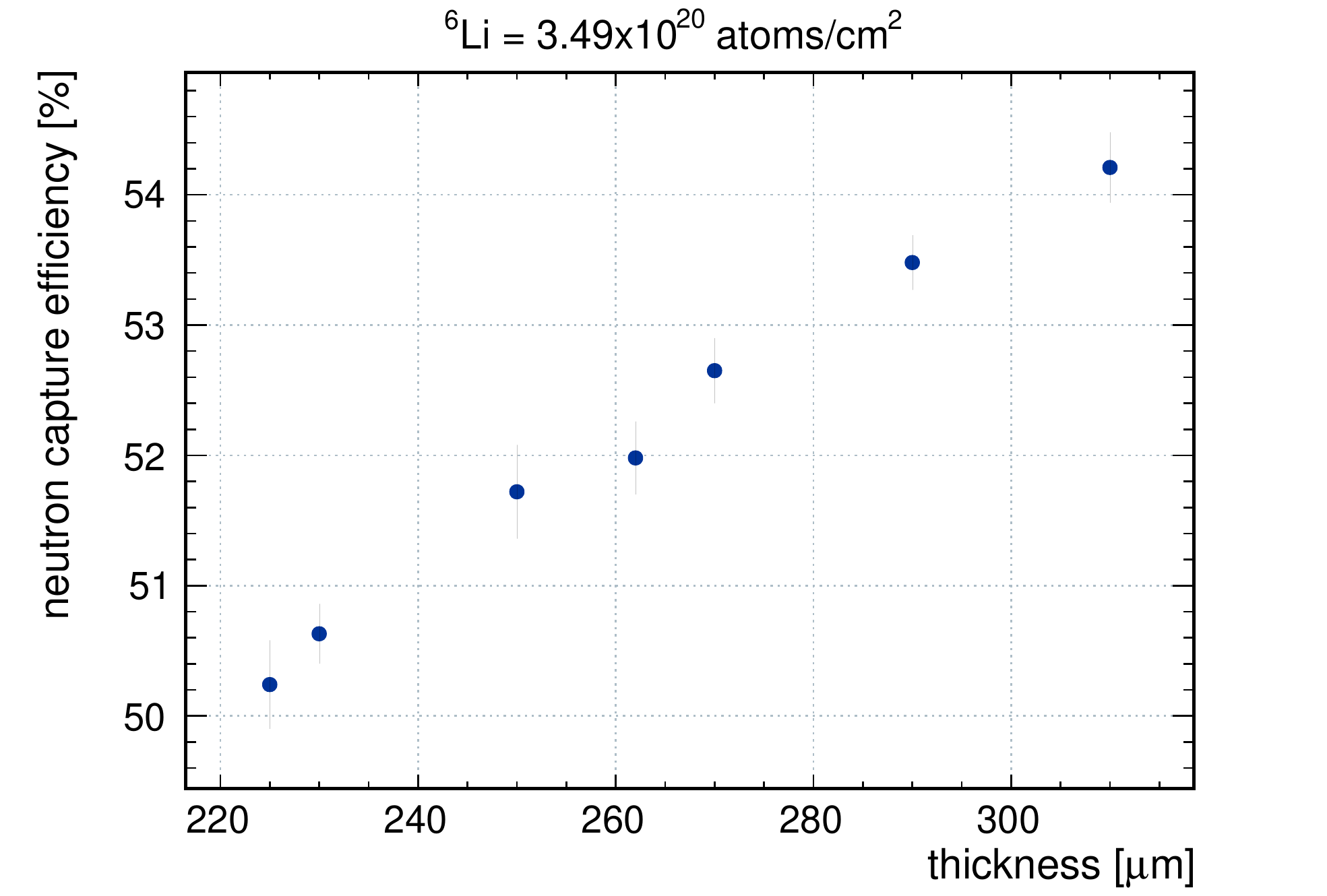}
\caption{\label{fig:neffvsthick} Simulated neutron capture efficiency as a function of the screen thickness for a fixed \lith~concentration of 3.49$\times$10$^{20}$ atoms/cm$^2$.}
\end{center}
\end{figure}

\begin{figure}[ht]
\begin{center}
\includegraphics[width=0.5\textwidth]{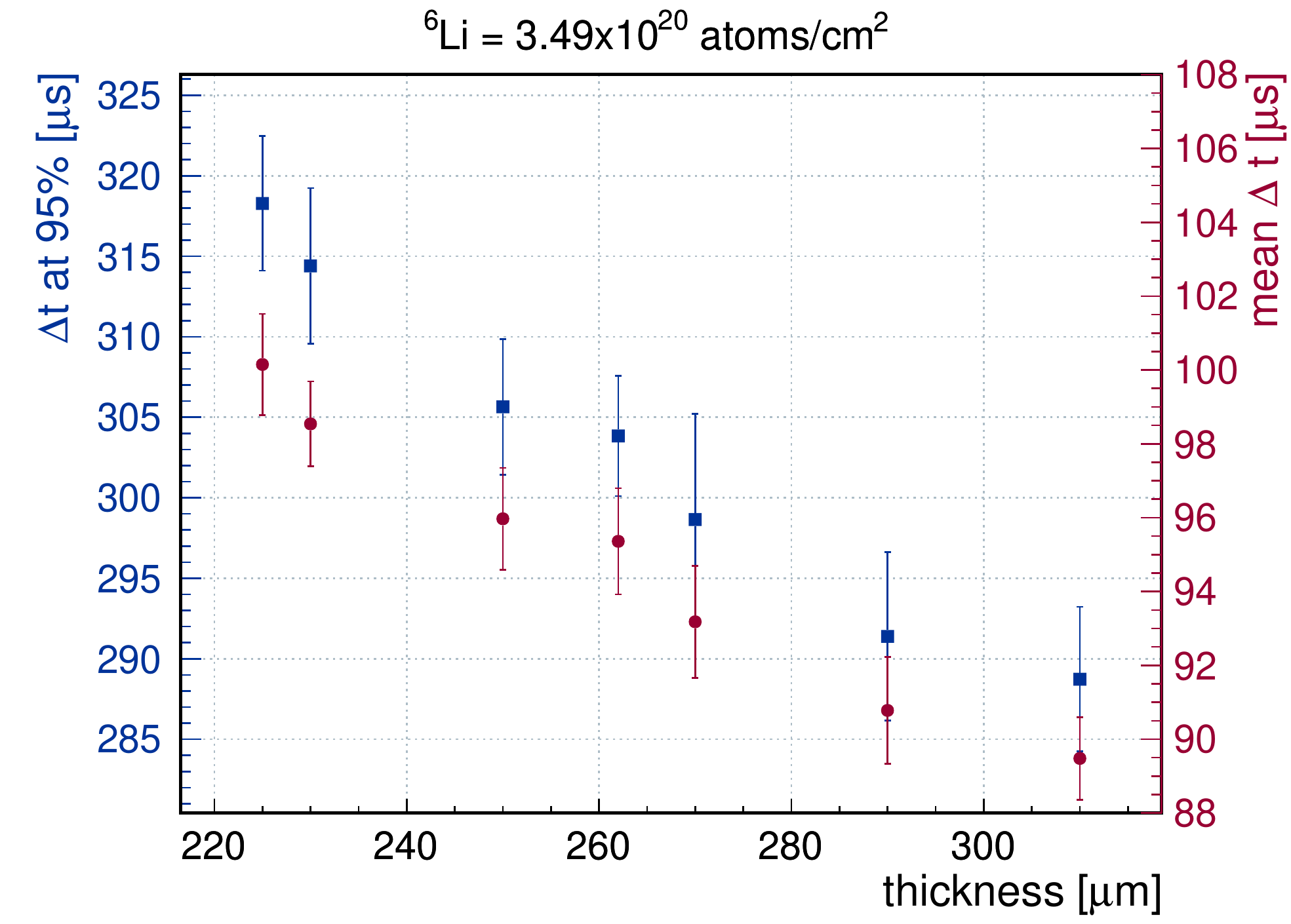}
\caption{\label{fig:dtvsthick} Simulated neutron capture time as a function of the screen thickness for a fixed \lith~concentration of 3.49$\times$10$^{20}$ atoms/cm$^2$. The capture time can be  expressed as the mean of the distribution of capture times (red) and as the value below which 95\% of all the capture times are contained (blue).}
\end{center}
\end{figure}

\begin{figure}[ht]
\begin{center}
\includegraphics[width=0.5\textwidth]{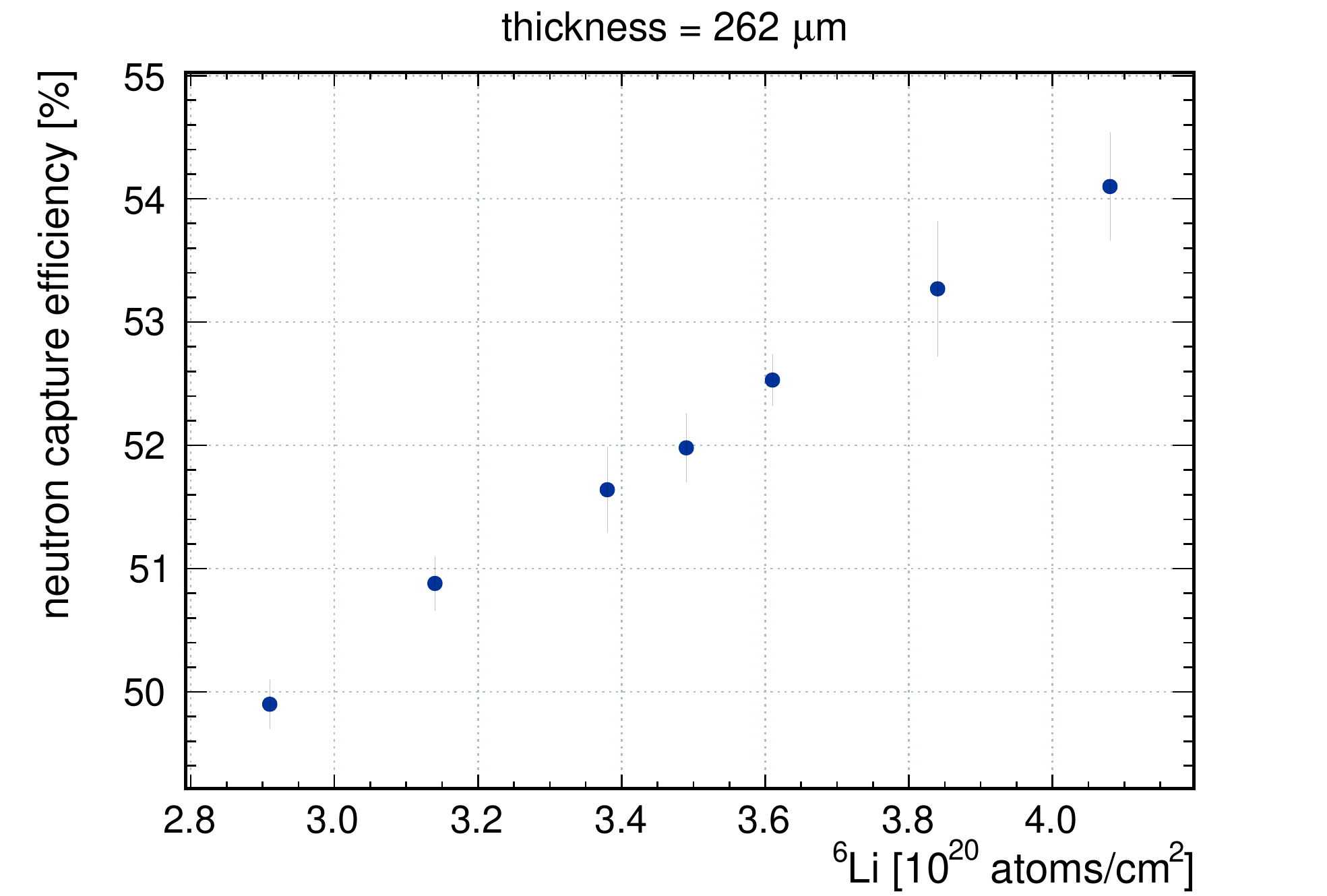}
\caption{\label{fig:neffvscon}  Simulated efficiency of neutron capture as a function of \lith~content. The thickness of the neutron screen is kept constant at a value of 262 microns.}
\end{center}
\end{figure}

\begin{figure}[ht]
\begin{center}
\includegraphics[width=0.5\textwidth]{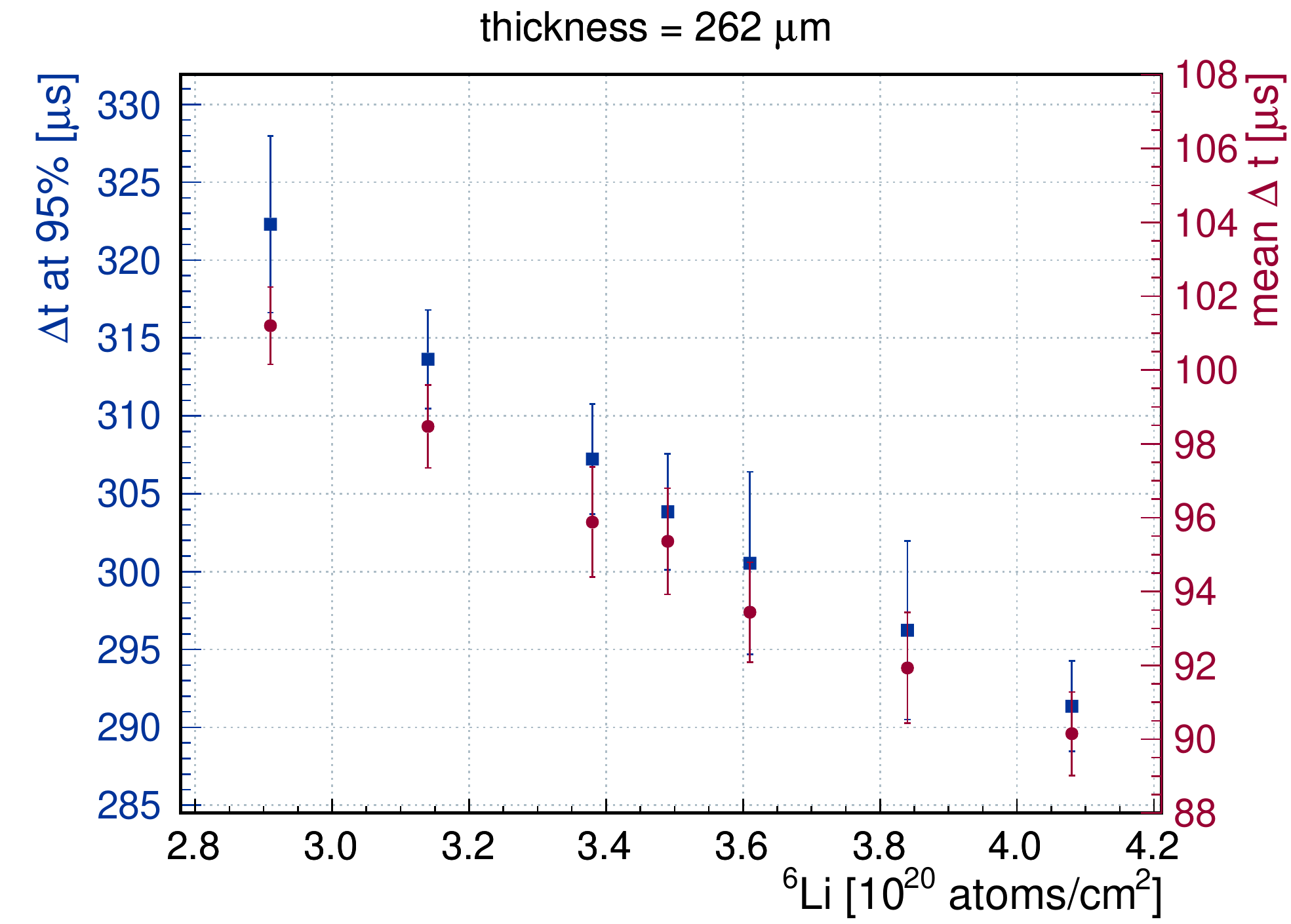}
\caption{\label{fig:dtvscon}  Simulated neutron capture time as a function of the \lith~content. The thickness of the neutron screen is kept constant at a value of 262 micron. The capture time is expressed in the same way as in Fig.~\ref{fig:dtvsthick}.}
\end{center}
\end{figure}

The dependence of the neutron capture efficiency and the time of capture as a function of screen thickness for a fixed \lith ~concentration are shown in Figs.~\ref{fig:neffvsthick}~and~\ref{fig:dtvsthick} respectively for the nominal value of \lith~concentration according to the manufacturer. The same quantities, as a function of \lith ~concentration for a fixed screen thickness, are shown in figs.~\ref{fig:neffvscon} and ~\ref{fig:dtvscon}. In each case, the dependence is compatible with a linear one and this can be used to convert the uncertainty in manufacturing parameters into uncertainties on the capture efficiency and capture time. A 10\% variation in concentration and thickness, compatible with tolerances provided by the screen manufacturer, results in a 1\% and 4\% variation in capture efficiency and capture time, respectively. The small dependence on the \lith\  content and screen thickness means that a tonne-scale detector with very high neutron capture uniformity can be realistically built using this technique.   

\subsection{Neutron light yield}
\label{n_coll}

The neutron detection efficiency depends on the light collection scheme adopted in the detector. If the number of photons detected is low, it is challenging to implement an efficient pulse-shape-based neutron trigger. Despite a large scintillation yield of up to 175\,k optical photons~\cite{Gnezdilov:2015pne} per neutron capture, a substantial fraction of these photons will be lost due to attenuation in the screen, due to internal reflection losses and absorption in the PVT and due to light losses in the optical fibre. Typical waveforms like the one presented in Fig.~\ref{fig:waveformNeutron} indicate that around a hundred photons distributed over tens of microseconds will reach the photon detector. This results in large statistical fluctuations that complicates detection by conventional threshold methods.

\begin{figure}[ht]
\begin{center}
\includegraphics[width=.5\textwidth]{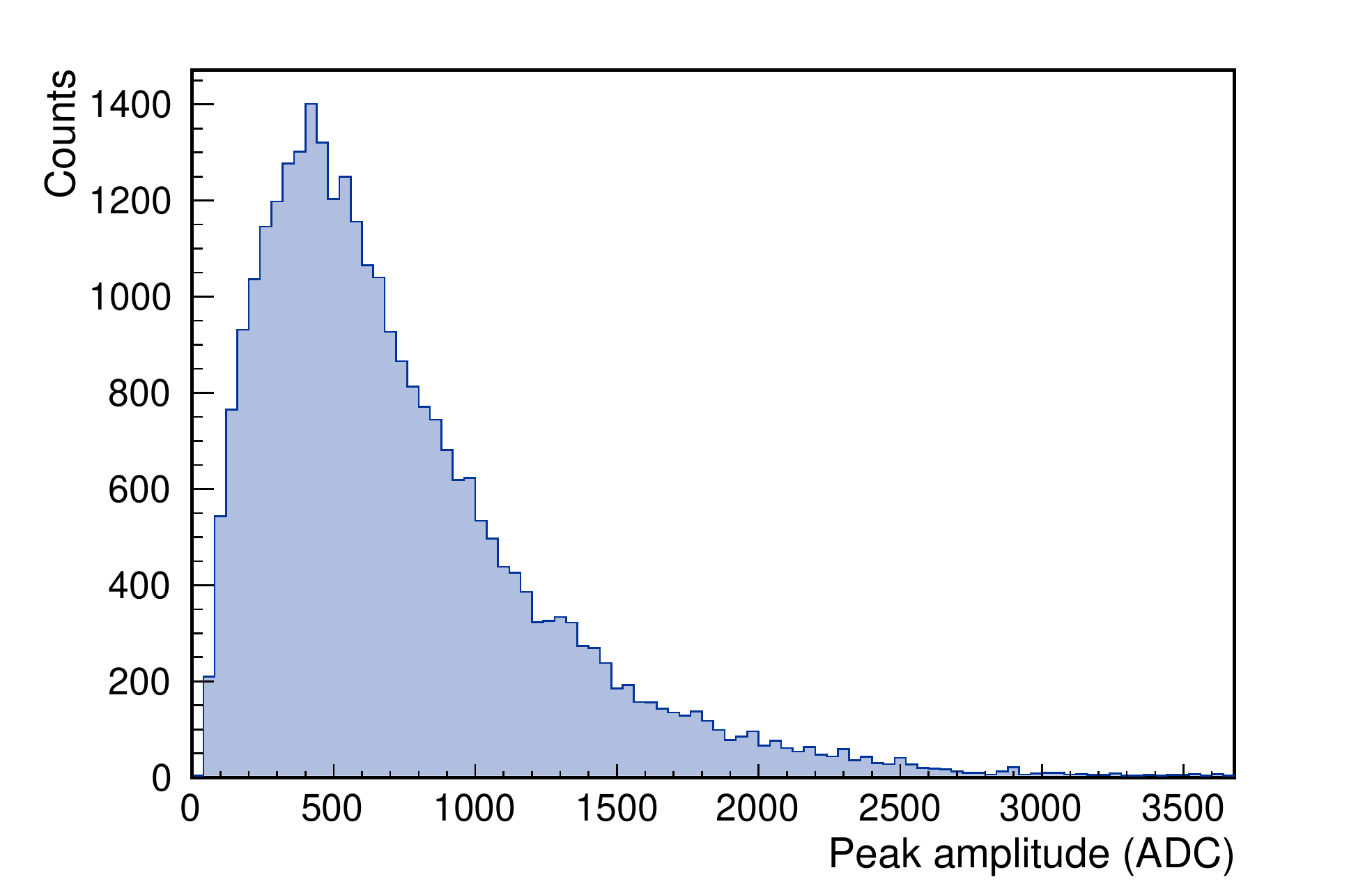}
\caption{Peak amplitude distribution of neutron signals produced by a \lith F:ZnS(Ag) screen recorded with a PMT, shown in ADC counts. \label{fig:neutronAmplitude}}
\end{center}
\end{figure}

\begin{figure}[ht]
\begin{center}
\includegraphics[width=.7\textwidth]{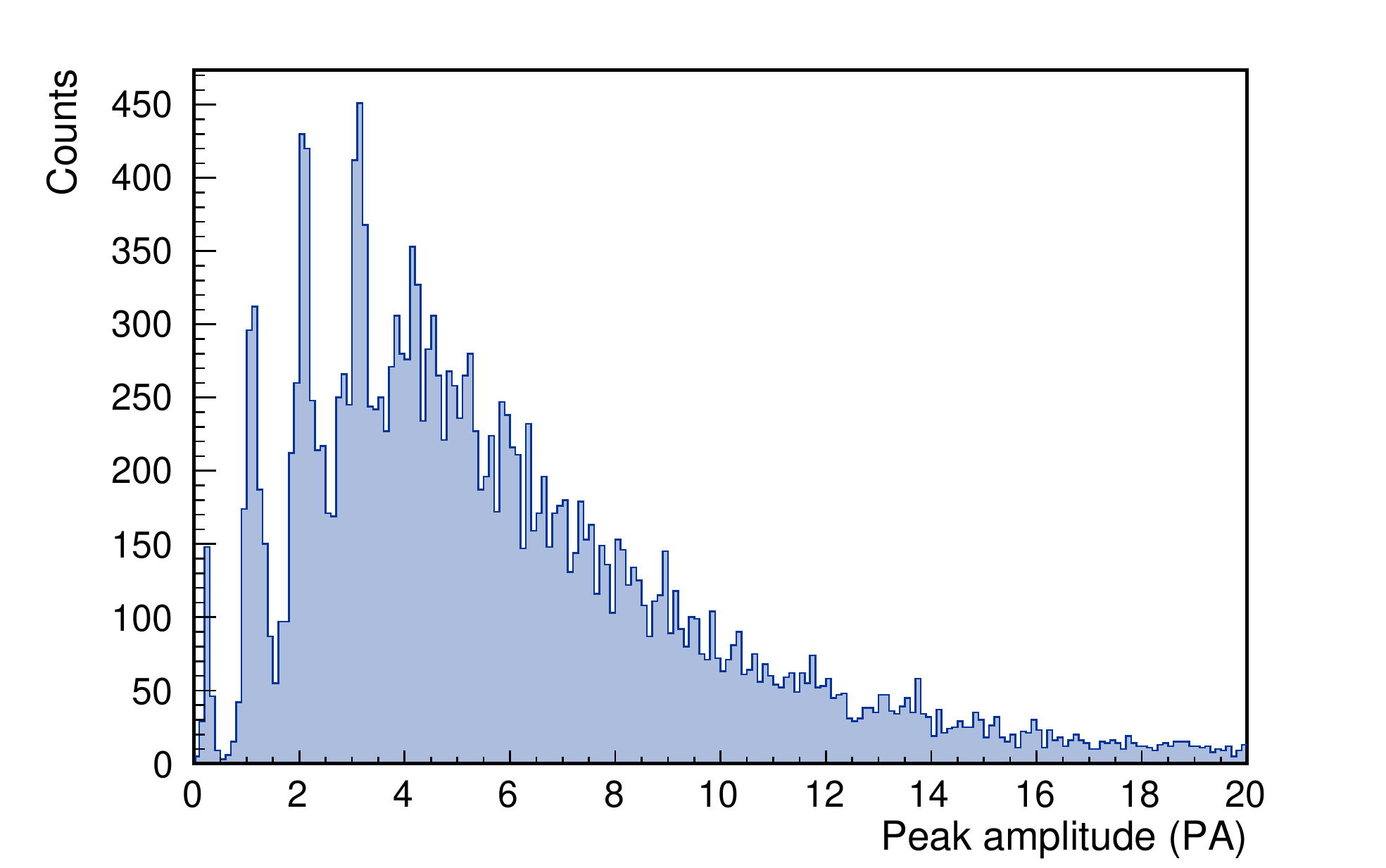}
\caption{Maximum amplitude distribution of the SiPM neutron signals in coincidence with the PMT. \label{fig:MPPCnAmplitude}}
\end{center}
\end{figure}

Fig.~\ref{fig:neutronAmplitude} shows the distribution of neutron peak amplitudes, measured with a 0.75\,inch Philips XP1911/UV photomultiplier tube directly coupled to a PVT cell, on the opposite face to a 250\,$\upmu$m thickness neutron screen. The signal was shaped and amplified with an Ortec 474 spectroscopic amplifier (no differentiation, integration time constant 500\,ns) and was digitised with a CAEN v1720 analogue to digital converter. The PMT signal was used to trigger recording of the light collected by a 8\,cm long WLS fibre with an attached SiPM. When triggering with a very low PMT threshold, a nearly unbiased neutron amplitude spectrum can be obtained for the SiPMs and is shown in Fig.~\ref{fig:MPPCnAmplitude}. The SiPM amplitude spectrum peaks at 4 to 5\, pixel avalanches (PA), with a substantial fraction of events below 3\,PA amplitude. 

A trigger scheme based on a simple amplitude threshold is difficult to implement in practice, especially with a SiPM dark noise rate that can be as high as hundred of kHz at a low pixel avalanche threshold. Therefore, a triggering method that exploits the full structure of the pulse is required for an efficient but sufficiently selective neutron trigger.

\section{Energy measurement}
\label{e_meas}

\subsection{Detector light yield}
\label{ly}

\begin{figure}[ht]
\begin{center}
\includegraphics[width=.5\textwidth]{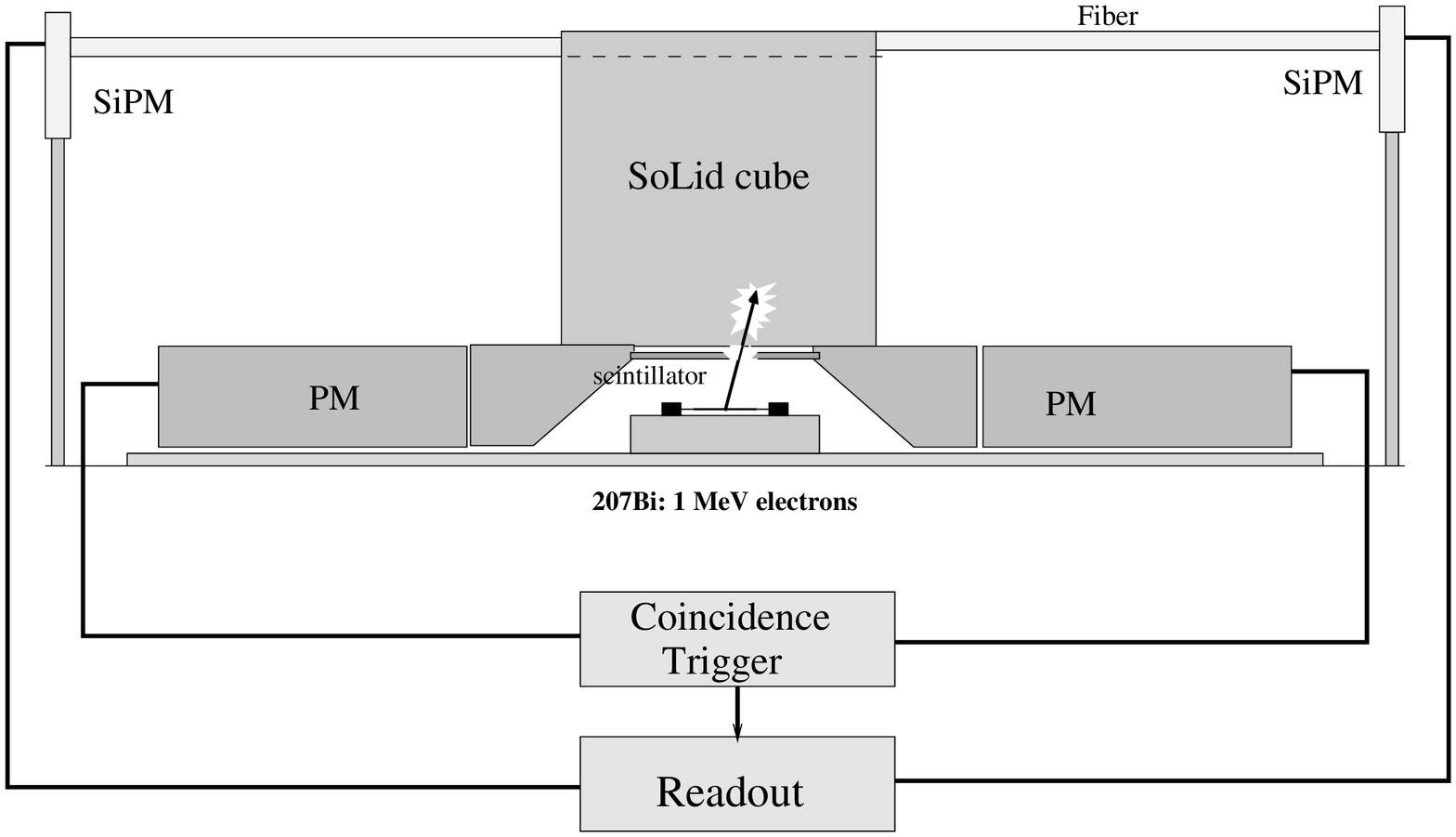}
\caption{\label{fig:LYsetup} Experimental setup to measure the detector cell light yield.}
\end{center}
\end{figure}

\begin{figure}[ht]
\begin{center}
\includegraphics[width=.5\textwidth]{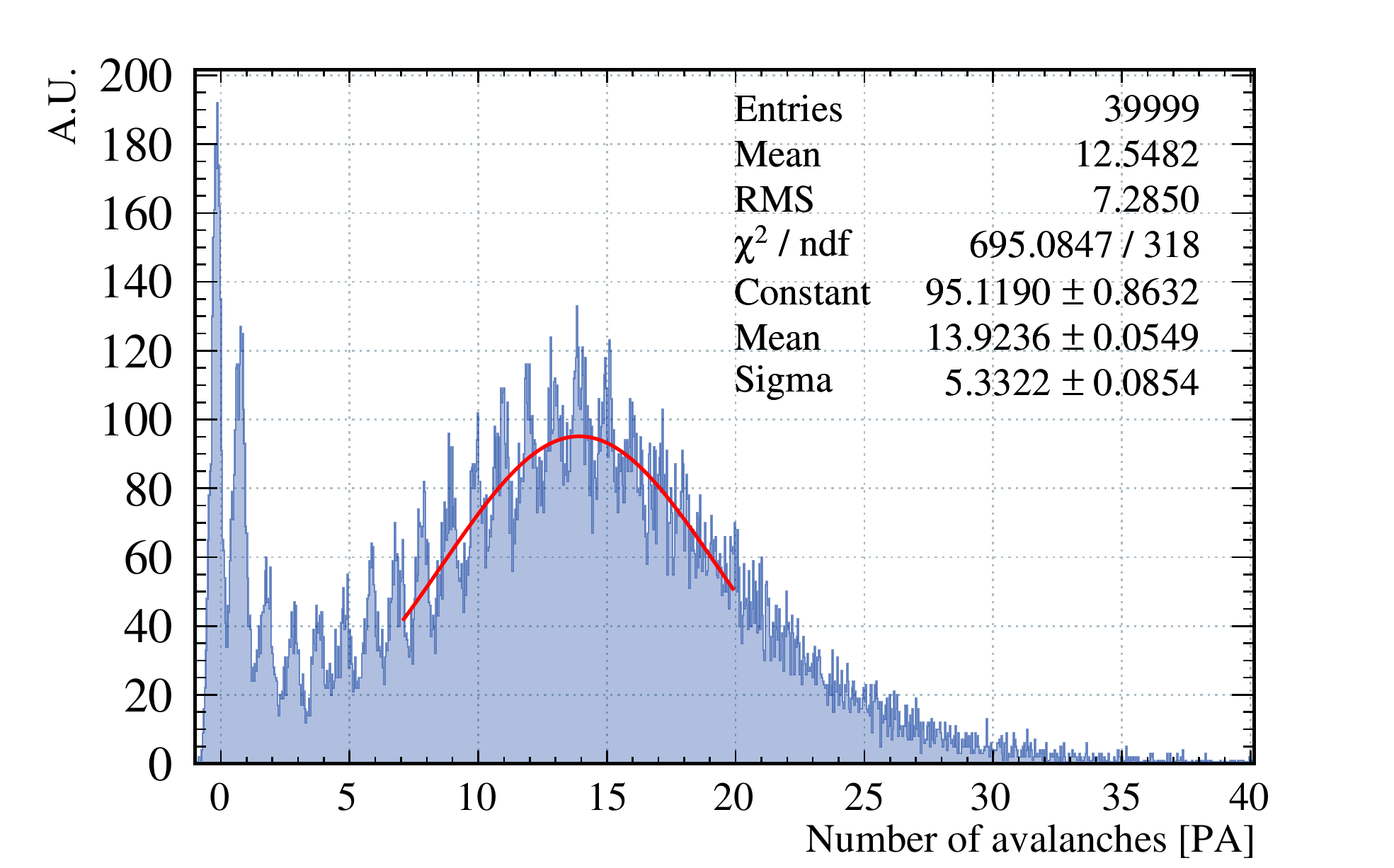}
\caption{\label{fig:LYspectrum} SiPM signal response to a $^{207}$Bi electron signal.}
\end{center}
\end{figure}

An accurate measurement of the energy in IBD interactions requires good light yield and uniformity of detector response. Achieving high light yield requires care in light collection, as the geometric acceptance of the fibre is small and the probability of photon capture in the fibre is low (5\% for double-clad fibres). It is therefore important to maximise reflectivity in the PVT cube and to optimise the coupling between optical components. A test bench has been set up to measure the light yield for various reflective materials and different types of surface treatment of the cubes and to compare different types of fibres and mirroring techniques. The setup is illustrated in Fig.~\ref{fig:LYsetup}. A $^{207}$Bi source is used to provide a mono-energetic electron signal from its 1064\,keV conversion electron decay, detected by a thin scintillator read out by two PMTs. The PMT coincidence signal is then used as a trigger logic to start the acquisition of SiPM signals at each end of a fibre. The measured pulse height spectrum is shown in Fig.~\ref{fig:LYspectrum} with a clear maximum at 14 PA, corresponding to the electron energy.

\begin{figure}[ht]
\begin{center}
\includegraphics[width=.5\textwidth]{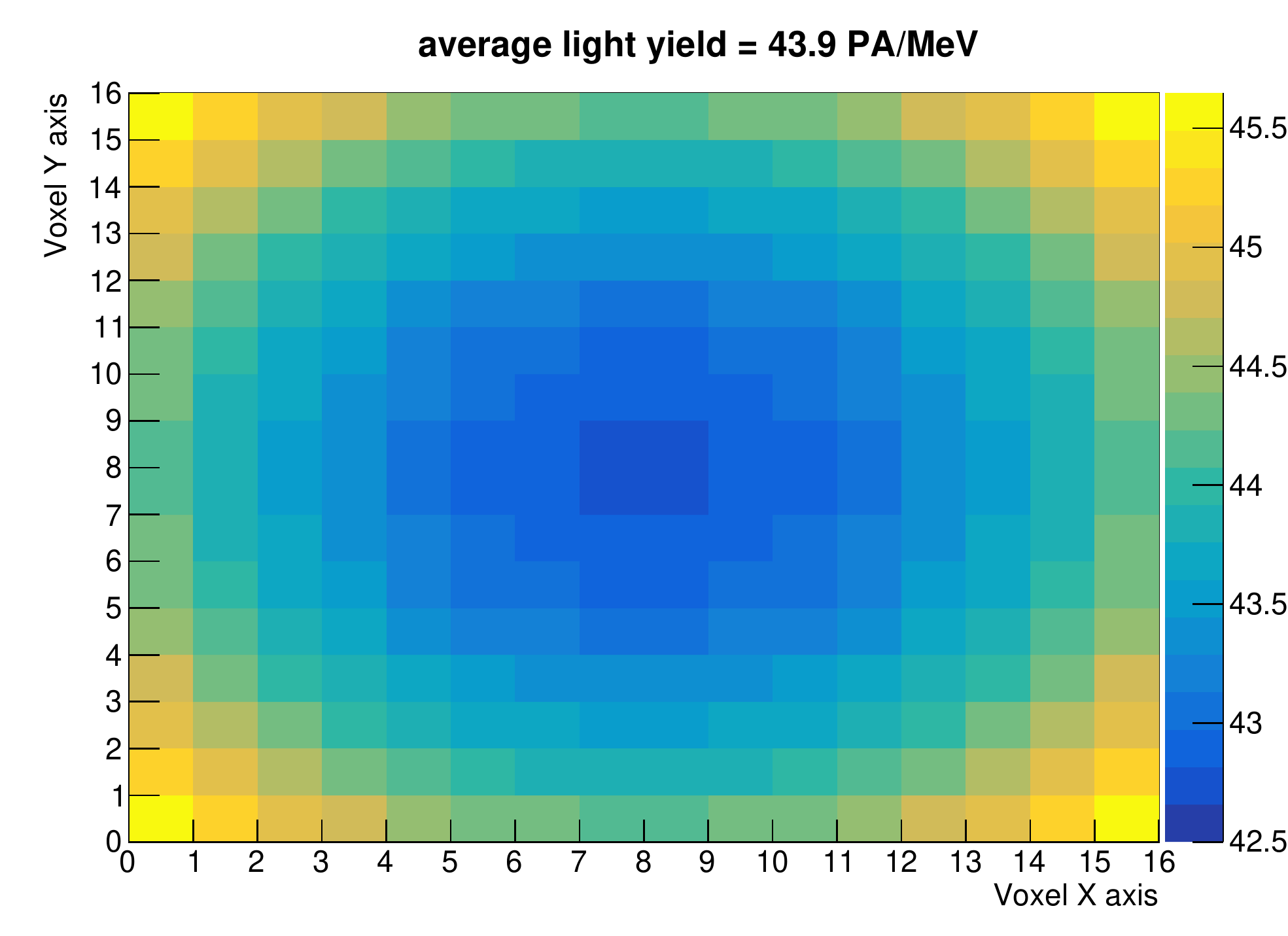}
\caption{\label{fig:LYuniformity} Light output per cube summed over fibres running along X and Y axes.}
\end{center}
\end{figure}

A different configuration, with four fibres coupled along two axes, with a SiPM at one end and a mirror at the other, gives a light yield of 40 PA/MeV from the sum of the fibre ends. This number can be used to estimate the energy resolution of the detector cell with an optimally adjusted SiPM operating voltage, of 14\%/$\sqrt{E\mathrm{(MeV)}}$. This resolution is dominated by statistical fluctuations due to the limited number of collected photons at low energies. This four-fibre setup produces a good uniformity of the light yield across each detector plane (Fig.~\ref{fig:LYuniformity}); the largest variation from the centre of the detector to the edge for a full-size detector is expected to be less than 10\%. 

\subsection{Energy reconstruction}
\label{e_rec}

\begin{figure}[ht]
\begin{center}
\includegraphics[width=.51\textwidth]{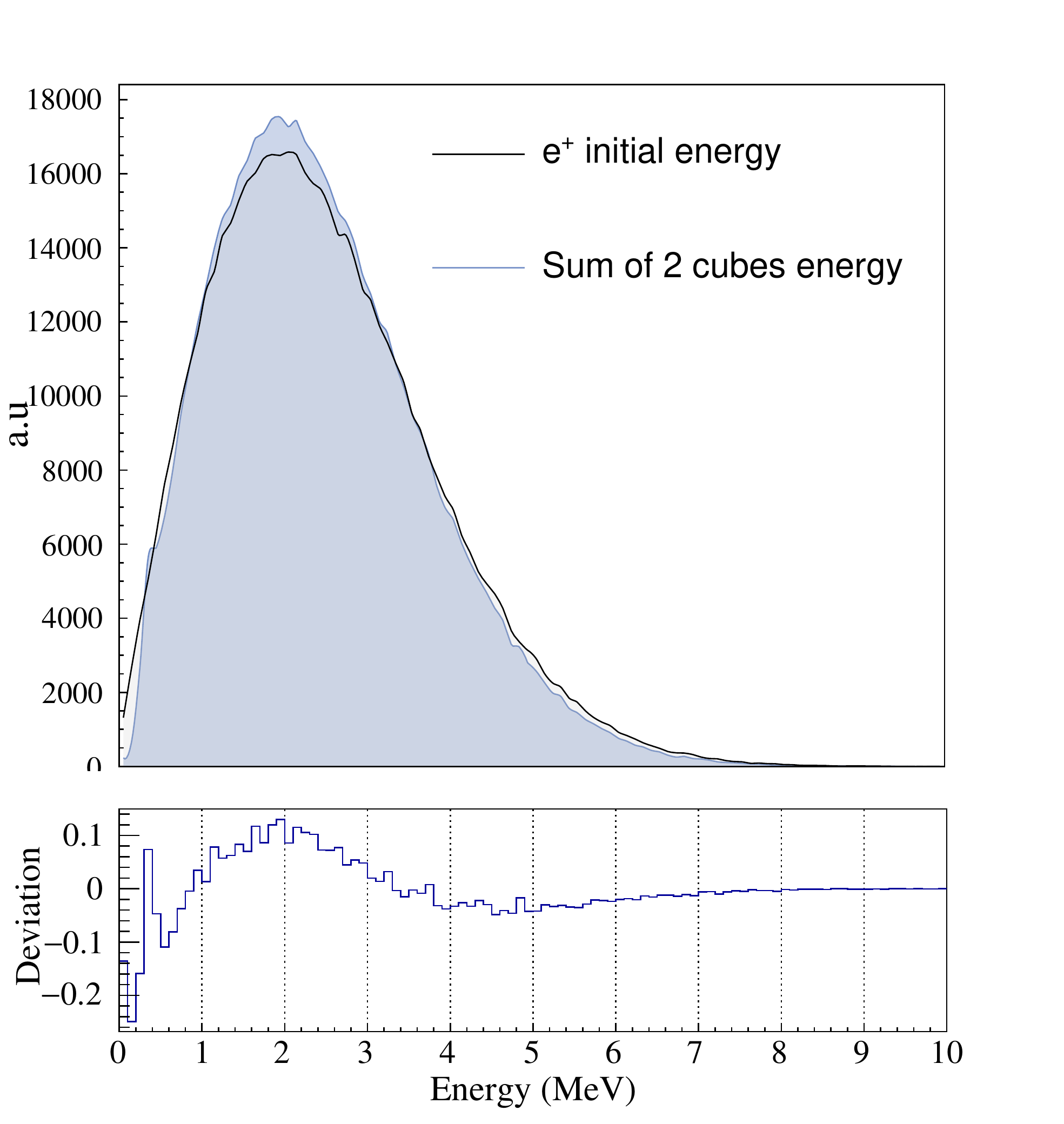}
\caption{\label{fig:ERec} Distribution of the positron initial energy (black line) from simulation and the reconstructed energy using the two voxels containing the highest energy deposit (blue histogram). The bottom figure shows the relative difference of the two distributions.}
\end{center}
\end{figure}

The \anu\ energy is estimated from the measurement of the positron energy in the detector. 
The positron annihilates and the visible energy is given by:
\begin{equation}
E_{\mathrm{vis}} = T_{\mathrm{e}} + 2M_{\mathrm{e}}, 
\end{equation}
where $T_{\mathrm{e}}$ is the positron kinetic energy and $2M_{\mathrm{e}}$ is the measured energy from the two 511~keV annihilation photons. In a large scintillator detector the whole
visible energy is used to estimate the positron energy. However, in the case of a segmented detector as proposed here, an alternative method can be used to estimate the positron energy. The deposits due to the annihilation photons may be excluded through a spatial cut and the visible energy (corresponding to the positron kinetic energy alone) added to the positron mass:
\begin{equation}
E_{\mathrm{e}}= T_{\mathrm{e}} + M_{\mathrm{e}} + E_{\mathrm{corr}},
\label{eq:Ereco}
\end{equation}
where the correction $E_{\mathrm{corr}}$ is an average value taking into account the contamination from annihilation photon energy deposited in the vicinity of the positron and the loss of energy in dead materials between cubes. Its value can be estimated accurately with simulation and cross-checked with calibration data. The difference between the true positron energy and the quantity measured using this technique is shown in Fig.~\ref{fig:ERec}. When summing over only the two cells with the highest energy deposition, the correction $E_{\mathrm{corr}}$ becomes relatively small (below 10\%) across the relevant energy range. This result combined with the small correction to be applied on light attenuation in the fibre and a precise SiPM gain equalisation means that the energy reconstruction can be controlled at the level of a few percents and constitutes a small systematic error on the energy measurement.  

\begin{figure*}[ht]
\begin{center}
\includegraphics[width=1.\textwidth]{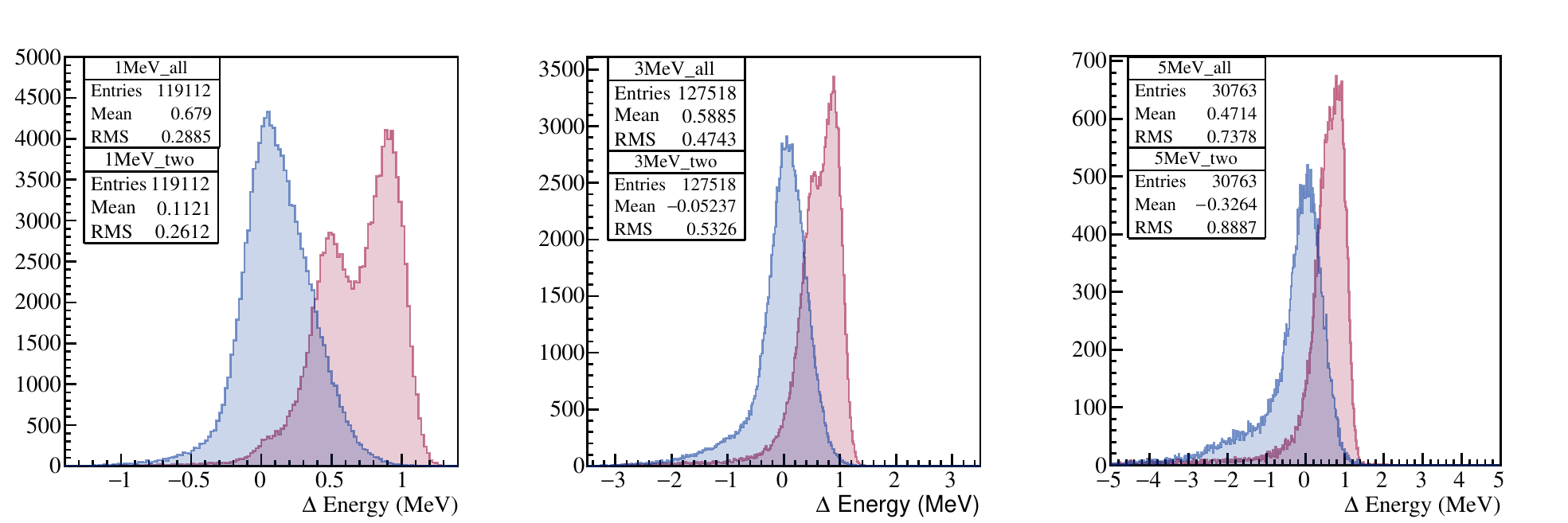}
\caption{The residual energy correction $E_{\mathrm{corr}}$ for positrons of kinetic energy (1, 3, 5)\,MeV. The red histograms show the correction to the visible energy obtained by integrating over all detector voxels in a 1m$^3$ detector and the blue histograms show the corrections when the visible energy is reconstructed by summing only the two most energetic deposits in the detector.  \label{fig:EresComp}  }
\end{center}
\end{figure*}

This strategy also reduces the effect of energy spread on the statistical component of the energy resolution. A coarser segmentation would result in more \gray~ energy being included in the energy estimator adding a source of significant variation to the measurement. Fig.~\ref{fig:EresComp} illustrates this point in the extreme case of summing the energy of mono-energetic positrons over all voxels (red distributions) and adding a Gaussian energy smearing of 6\%/$\sqrt{E\mathrm{(MeV)}}$, typical of large volume detectors. These results are compared to the estimator introduced in this section. Despite a larger energy smearing for the proposed estimator, the energy estimator is less biased and the distributions have a comparable width. This result shows that, in the case of a relatively small segmented detector, the energy reconstruction technique is as important as the intrinsic per-channel energy resolution.     

\section{Antineutrino detection}

In order to evaluate the performance of the proposed SoLid detector and to provide objective guidance for the optimisation of detector, trigger and reconstruction, a preliminary set of IBD selection criteria have been established based on simulation studies.

A G{\sc eant4} simulation of a cubic metre detector was performed, with \anu\ interactions generated uniformly in the detector's PVT volume. Signal selection criteria are based on the time difference between positron and neutron signals, $\Delta t$, the energy of the positron signal, the distance between neutron and positron cell, $\Delta R_\mathrm{ne}$ and a measure of the spatial extent of the prompt signal energy deposit, $\Delta R_\mathrm{ee}$.

In reactor experiments, two type of backgrounds are commonly encountered:

\begin{enumerate}
\item Accidental background, due to time coincidence between a prompt signal (\gray) and a  neutron signal close enough in time to be associated. The two signals are unconnected in origin and for instance could originate from fission processes in the reactor, from activation of material in the reactor hall, or from natural radioactivity present in the building.
\item Correlated background, due to a prompt and neutron signal originating from a common source, such as nuclear decay of radioactive elements or a fast neutron entering the detector.  
\end{enumerate}

Since the SoLid detector operates without significant overburden, neutrons induced by cosmic rays in the atmosphere are used as a representative correlated background. They are simulated using a realistic spectrum~\cite{Gordon2004} with a capture rate of a few Hz in a one-tonne detector. The \gray~ energy spectrum used to calculate the random background is proportional to $1/E^2$ and the total rate of \gray~ above 1\,MeV entering the detector is 10 kHz. The coincidence time window is 250\,$\upmu$s. The simulation uses a configuration of two \lith F:ZnS(Ag) screens per cube and the detector is surrounded by a 5\,cm polyethylene moderator, resulting in a neutron capture efficiency of 71\%. No additional shielding is present around the detector. 

A simple selection analysis was performed for IBD events and the two classes of background. The details of the selection cuts and the survival probabilities of each category are shown in Table~\ref{tab:cuts}.    

\begin{table}[htp]
\caption{Cumulative survival probability obtained from simulated IBD events and two main backgrounds. 
}
\begin{center}
\begin{tabular}{c  c  c  c }
\hline 
Cut & \multicolumn{3}{c}{Survival probability (\%)} \\
{}  &  IBD & Cosmic n & Random \\
\hline 
\hline 
$E_\mathrm{em} \geq1.0 \text{MeV} $                 & 46   & 15   &  12 \\ 
$\min(\Delta R_\mathrm{ne}) \leq 3 \text{ cubes}$   & 44  & 10   &  0.32 \\
$\max(\Delta R_\mathrm{ee}) \leq 1 \text{ cube}$   & 42  & 2.8  & 0.23 \\
\hline 
\end{tabular}
\end{center}
\label{tab:cuts}
\end{table}%

The optimisation of cut values is made assuming a neutron efficiency of 54\% for IBD events,  taking into account a neutron detection efficiency of 0.80, expected to be achievable with an appropriate trigger strategy. The neutron efficiency for background events is 24\% and lower than signal events due to the wider range of atmospheric neutron kinetic energy. This study indicates that analysis cuts that combine precise spatial and timing information can reduce substantially both categories of backgrounds by roughly a factor of a hundred for accidental background and a factor of ten for correlated background.    

For a detector like the one used in SoLid, located at a distance of 6.5 m of the BR2 reactor operated at 60 MW thermal power, around 630 IBD events can be detected per day ($7.3 \times 10^{-3}$ Hz) after all cuts. The accidental background rate in this case is 5 Hz and the rate of correlated background is 0.48 Hz based on the Monte Carlo efficiency. After all cuts are applied the combined background rate is 0.09 Hz corresponding to a S:B of 1:10. This is a very significant reduction achieved through the use of the segmentation. Unfortunately lowering the background by another order of magnitude is only possible with passive neutron shielding and all experiments operating close to the surface have to use both ways of mitigating the background.

\section{Conclusions}
\label{conc}
\sloppy
We have presented a design for a new compact composite scintillator detector based on \LiF\ and PVT intended for measuring IBD interactions with percent level precision. The detector makes efficient use of the total available target volume by detecting IBD positron and neutron close to the interaction point, providing both good position resolution and large background rejection. Simulation studies and experimental tests have shown that this design provides a high neutron capture efficiency of up to 73\%, an IBD efficiency of 42\% and a light yield that translates into an intrinsic energy resolution of 14\%/$\sqrt{E\mathrm{(MeV)}}$. It was also shown that a large scale system can be built that has very with good uniformity of response using readily available scintillator products. Despite the small size of the detector, an energy estimator that does not degrade the energy resolution due to reconstruction can be implemented with the chosen segmentation. This fine granularity also provides a new way of rejecting the main IBD backgrounds, which combined with passive shielding, can provide a S:B of 1:1 close to the BR2 reactor core.  

\newpage

\section{Aknowledgments}
This work was supported by the following funding agencies: Agence Nationale de la Recherche grant ANR-$16­\mathrm{CE}31­0018­03$, Institut Carnot Mines, CNRS/IN2P3 et Region Pays de Loire, France; FWO-Vlaanderen and the Vlaamse Herculesstichting, Belgium; The U.K. groups acknowledge the support of the Science \& Technology Facilities Council (STFC), United Kingdom; We are grateful for the early support given by the sub-department of Particle Physics at Oxford and High Energy Physics at Imperial College London. We thank also our colleagues, the administrative and technical staffs of the SCK\raisebox{-0.8ex}{\scalebox{2.8}{$\cdot$}}CEN for their invaluable support for this project.  Individuals have received support from the FWO-Vlaanderen and the Belgian Federal Science Policy Office (BelSpo) under the IUAP network programme; The STFC Rutherford Fellowship program and the European Research Council under the European Union's Horizon 2020 Programme (H2020-CoG)/ERC Grant Agreement \mbox{n. 682474} (corresponding author); Merton College Oxford.





\bibliographystyle{model1a-num-names}
\bibliography{SoLid_tech_paper.bib}







\end{document}